\documentclass[abbrv,aps,prb,twocolumn,showpacs,preprintnumbers,amsmath,amssymb,
superscriptaddress,nofootinbib]{revtex4}

\usepackage{graphicx}
\usepackage{dcolumn}
\usepackage{bm}

\begin{document}
\title{Local correlations and hole doping in NiO}
\author{J. Kune\v{s}}
\affiliation{Theoretical Physics III, Center for Electronic Correlations and Magnetism, Institute of Physics, 
University of Augsburg, Augsburg 
86135, Germany}
\email{jan.kunes@physik.uni-augsburg.de}
\affiliation{Institute of Physics,
Academy of Sciences of the Czech Republic, Cukrovarnick\'a 10,
162 53 Praha 6, Czech Republic}
\author{V.~I.~Anisimov}
\affiliation{Institute of Metal Physics, Russian Academy of
Sciences-Ural Division, 620041 Yekaterinburg GSP-170, Russia}
\author{A.~V.~Lukoyanov}
\affiliation{Ural State Technical University-UPI,
620002 Yekaterinburg, Russia}
\author{D. Vollhardt}
\affiliation{Theoretical Physics III, Center for Electronic Correlations and Magnetism, Institute of Physics,
University of Augsburg, Augsburg
86135, Germany}
\date{\today}

\begin{abstract}
Using a combination of {\it ab initio} bandstructure methods and dynamical mean-field theory
we study the single-particle spectrum of the prototypical charge-transfer insulator
NiO. Good agreement with photoemission and inverse-photoemission spectra
is obtained for both stoichiometric and hole-doped systems.
In spite of a large Ni-$d$ spectral weight at the top of the valence band the doped holes are found 
to occupy mainly the ligand $p$ orbitals. Moreover, high hole doping
leads to a significant reconstruction of the single-particle spectrum accompanied
by a filling of the correlation gap.
\end{abstract}
\maketitle

\section{Introduction}
Already in 1937, at the outset of modern solid state physics, de Boer and Verwey \cite{deboer} drew attention
to the surprising properties of materials with incompletely filled 3$d$-bands, such as NiO. This
observation prompted Peierls and Mott \cite{mott} to discuss the interaction between the electrons.
Ever since transition metal oxides (TMOs) were investigated intensively. 
This interest further increased when it was discovered that
TMOs display an amazing multitude of ordering and electron correlation phenomena, including 
high temperature superconductivity, colossal magnetoresistance and Mott metal-insulator transitions \cite{imada}.
In the late 1950's MnO and NiO were taken as the textbook examples of antiferromagnets. 
However, when the importance of local Coulomb correlations in the transition metal $d$-shell  
was realized TMOs were considered candidates for Mott insulators \cite{mott}. 
In the mid 1980's Zaanen, Sawatzky and Allen (ZSA) introduced their classification of 
TMOs and related compounds into Mott-Hubbard and charge-transfer (CT) systems \cite{zaa85}. 
In the early TMOs the ligand $p$-band is located well below the transition metal $d$-band 
and thus plays a minor role in the low energy dynamics. Such a case, called 
Mott-Hubbard system in the ZSA scheme, is well described by a multi-band Hubbard model.
On the other hand, the late TMOs belong to the CT type where the $p$-band is situated between
the interaction split $d$-bands. A more general Hamiltonian where  
the $p$-states are explicitly included is then needed, which can be viewed as a combination
of multi-band Hubbard and Anderson lattice models. 
A major impulse for detailed investigations of CT systems, and especially of their hole doped regime, 
came with the discovery of high temperature superconductivity in cuprate perovskites. 
While the standard three-band Hamiltonian for cuprates \cite{emery} contains only one $d$-orbital
per lattice site, the description of cubic transition metal monoxides, the prominent member of which
is NiO, requires the full set of $d$-orbitals. The latter are of interest not only
for fundamental research, but play an important role also in fields such as geophysics \cite{cohen}.
Furthermore, recent progress in high pressure experiments \cite{yoo}
made the insulator-to-metal transition in some TMOs accessible in the laboratory, 
providing yet another stimulus for theoretical investigations.

We report a computational study of NiO combining {\it ab initio}
band structure calculations in the local density approximation (LDA) with the dynamical 
mean-field theory (DMFT), an approach known as LDA+DMFT \cite{ldadmft}.
By treating the local correlations and the Ni $3d$ - O $2p$ hybridization on the same 
footing we provide a description of the full valence and conduction band spectra 
of a CT system with strong hybridization.
We will show that a good quantitative agreement with
photoemission and inverse-photoemission data can be obtained thereby. 
This provides a solid foundation for the
subsequent investigation of hole doping, a question of broader interest mainly in the context
of cuprates. It will be shown that the behavior of the doped holes 
clearly reveals the CT character of NiO.

NiO is a type II anti-ferromagnet ($T_N=523$ K) with
a magnetic moment of almost 2 $\mu_B$ and a large gap surviving well above $T_N$.
The standard LDA band theory predicts NiO to be a metal \cite{mat72}, 
or an antiferromagnetic insulator \cite{ter84} if spin polarization is allowed. 
A severe underestimation of the gap and the magnetic moment
suggests, however, that the Slater antiferromagnetic state 
obtained within LDA does not describe the true nature
of NiO. On the other hand exact diagonalization studies on small clusters 
were quite successful in describing the single- and two-particle 
spectra \cite{fuj84}, showing that the local Coulomb interactions are important. 
This made it clear that an explicit treatment of Coulomb interactions within the
$3d$ shell is needed, and methods such as
LDA+U \cite{ani91}, self-interaction correction \cite{sva91}, or GW \cite{ary95} were introduced.
The static, orbitally dependent self-energy of LDA+U 
enforces a separation of the occupied and unoccupied $d$-bands and thus opens a gap comparable
to experiment. This in turn leads to a significant improvement of the description of static properties
such as the local moment or the lattice dynamics \cite{savrasov}. 
However, the LDA+U method is limited to an ordered state and does not
yield the electronic excitations and the effect of doping correctly.  

A systematic inclusion of dynamical correlations was made possible by 
dynamical mean-field theory \cite{met89,rmp,pt}. 
Since its introduction DMFT proved to be a powerful
tool for the investigation of electronic systems with strong local correlations.
In connection with band-structure methods, the LDA+DMFT scheme \cite{ldadmft} provides access to material
specific single-particle spectra as well as more general correlation functions.
Applications of LDA+DMFT so far were mostly limited to Mott-Hubbard systems, where  
the ligand states are integrated out before the correlation problem is solved.
Recently, Ren {\it et al.} \cite{ren06} applied this approach to NiO and were able to
obtain a realistic gap and the near-gap spectra. However, this approach takes
into account only $d$-electrons, such that 
the orbital character of the valence and conduction band are bound to be the same, 
the high frequency incoherent features in the valence band are missing, and
the hole doping cannot be described realistically. 
In this work we go beyond such limitations by working with the ligand $p$ states explicitly.

 \begin{table}
 \caption{\label{tab:1} Orbital occupancies and the local moment on the Ni site for
  different hole dopings.}
 \begin{ruledtabular}
 \begin{tabular}{c|cccc}
  $n_h$ & $n_{e_g}$ & $n_{t_{2g}}$ & $n_p$ & $m_d$ \\
\hline
  0 & 0.547 & 1.000 & 0.969 & 1.85 \\
  0.6 & 0.531 & 0.994 & 0.885 & 1.61 \\
  1.2 & 0.530 & 0.980 & 0.800 & 1.45 \\
 \end{tabular}
 \end{ruledtabular}
 \end{table}
 \section{Computational details}
LDA+DMFT proceeds in two steps: (i) construction of the effective Hamiltonian
from converged LDA calculation, and (ii) solution of the corresponding DMFT equations.
Here we use the projection onto Wannier functions \cite{ani05} to obtain an eight-band $p-d$ Hamiltonian
\begin{equation}
\label{eq:ham}
\begin{split}
H=&\sum_{\mathbf{k},\sigma}\bigl(h_{\mathbf{k},\alpha\beta}^{dd}d_{\mathbf{k}\alpha\sigma}^{\dagger}
d_{\mathbf{k}\beta\sigma}+h_{\mathbf{k},\gamma\delta}^{pp}p_{\mathbf{k}\gamma\sigma}^{\dagger}
p_{\mathbf{k}\delta\sigma}+ \\& h_{\mathbf{k},\alpha\gamma}^{dp}d_{\mathbf{k}\alpha\sigma}^{\dagger}
p_{\mathbf{k}\gamma\sigma}+h_{\mathbf{k},\gamma\alpha}^{pd}p_{\mathbf{k}\gamma\sigma}^{\dagger}
d_{\mathbf{k}\alpha\sigma}\bigr)+\\&\sum_{i,\sigma,\sigma'}U_{\alpha\beta}^{\sigma\sigma'}n^d_{i\alpha\sigma}n^d_{i\beta\sigma'}.
\end{split}
\end{equation}
Here $d_{\mathbf{k}\alpha\sigma}$ and $p_{\mathbf{k}\gamma\sigma}$ 
are Fourier transforms of $d_{i\alpha\sigma}$ and $p_{i\gamma\sigma}$, which annihilate the $d$ or $p$ electron 
with orbital and spin indices $\alpha\sigma$ or $\gamma\sigma$ in the $i$th unit cell, 
and $n^d_{i\alpha\sigma}$ is the corresponding occupation number operator.
The elements of $U_{\alpha\beta}^{\sigma\sigma'}$ matrix are parameterized
by $U$ and $J$. The constrained LDA calculation yields $U$=8 eV and $J$=1 eV \cite{ani91}.
To account for the Coulomb interaction already present in LDA
we renormalize the $dd$-diagonal elements of the LDA Hamiltonian by the double counting correction
\begin{equation}
h^{dd}_{\mathbf{k},\alpha\beta}=\Tilde{h}^{dd}_{\mathbf{k},\alpha\beta}(\mathbf{k})-(N_{\text{orb}}-1)
\bar{U}n_{LDA}\delta_{\alpha\beta}
\end{equation}
where $n_{LDA}$ is the average LDA occupation per orbital and $N_{\text{orb}}$=10 is the total number of 
orbitals within the shell.
\begin{figure}
\includegraphics[width=\columnwidth,clip]{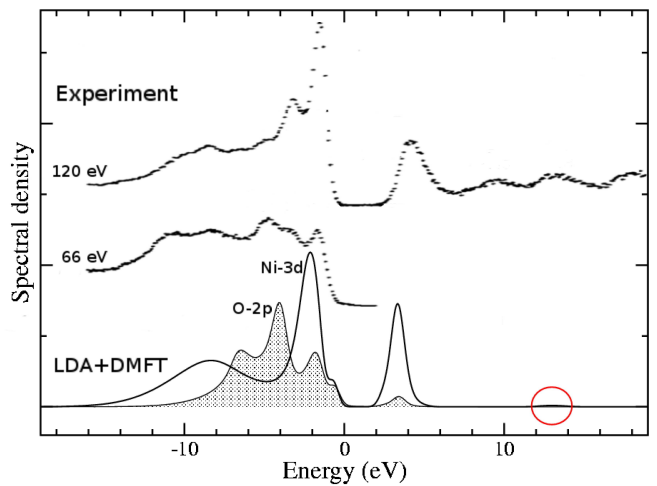}
\caption{\label{fig:total}(color online)  Theoretical Ni-$d$ (solid line) and O-$p$ (shaded) resolved spectral densities compared
to photoemission and inverse photoemission data obtained at
120 eV and 66 eV photon frequencies after Ref. \onlinecite{saw84}. Gaussian broadening of 0.6 eV full width at
half maximum corresponding to the experimental resolution was applied to the theoretical curves.
The circle marks position of the $d^{10}\underline{L}$ excitation.}
\end{figure}

Next we iteratively solve the DMFT equations on the Matsubara
contour, a key part of which is the auxiliary impurity problem solved by quantum Monte-Carlo (QMC) method \cite{qmc}.  
The results reported here were obtained at T=1160 K.
To obtain the single-particle spectral functions analytic continuation to real frequencies 
is performed using the maximum entropy method \cite{mem}. 
Applying QMC to a gapped system requires careful assessment of ergodicity and autocorrelation issues \cite{fw}. 
\begin{figure}
\includegraphics[angle=270,width=\columnwidth,clip]{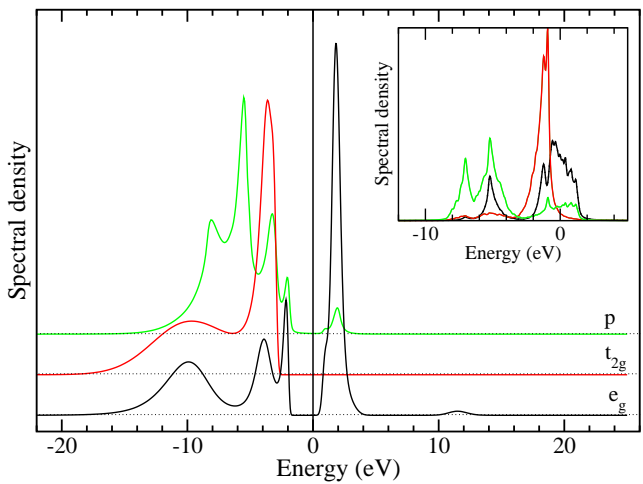}
\caption{\label{fig:resolved}(color online) Orbitally resolved O $2p$ and Ni $3d$ spectral densities  
(offset for better resolution).
Substantial spectral weight transfer relative to the LDA results (see inset) 
is observed.} 
\end{figure}

\section{Results and discussion}
\subsection{Single particle spectra}
The orbital occupations, shown in Table \ref{tab:1}, and the local moment of 1.85 $\mu_B$ 
obtained from the paramagnetic DMFT solution correspond to a $d^8$ groundstate of the Ni ion
with two ferromagnetically coupled holes of $e_g$ symmetry. 
In Fig. \ref{fig:total} the calculated spectral densities resolved into Ni $3d$
and O $2p$ contributions
are compared to photoemission and inverse photoemission data \cite{saw84}. 
Using the full $p-d$ Hamiltonian we are able to cover the entire 
valence and conduction bands spectra. Features corresponding to $4s$ and $4p$
bands at 10 eV and 13 eV, respectively, are not included in the theoretical spectrum. 
As shown by Eastman and Freeouf \cite{eastman} 
the relative intensity of the $2p$ contribution increases with decreasing photon energy.
Therefore the 120 eV spectrum is dominated by Ni $3d$ emission, while at 66 eV photon energy
the O $2p$ contribution peaked around -4 eV is resolved (for a detailed
orbital decomposition see Ref. \cite{eastman}). 
The theoretical spectrum very well reproduces the experimental features, including
the size of the gap, the $d$ character of the conduction band, the broad $d$ peak at -9 eV,
the position of the $p$-band, and the strong $d$ contribution at the top 
of the valence band. 
While the gap and the Hubbard subbands can be described already with the 
static theory (LDA+U) \cite{ani91}, a dynamical treatment is apparently needed
to capture the substantial redistribution of spectral weight between the incoherent (-9 eV)
and resonant (-2 eV) features in the $d$ spectrum. 
For a more detailed analysis we show in Fig. \ref{fig:resolved} 
the spectral density resolved into $e_g$ and $t_{2g}$ representations.
In agreement with other studies \cite{ren06,fuj84}
we find the conduction band to have a pure $e_g$ character.
The dominant feature of the valence spectrum is a distribution of spectral
weight between the broad peak at high frequency and sharp peak(s) at the gap edge.
The origin of these structures was discussed by Fujimori {\it et al.} \cite{fuj84} in terms 
of eigenstates of a NiO$_6$ cluster. Emitting a $d$ electron from the
$d^8$ state the system can end up either directly in a $d^7$ final state (broad peak)
or in a $d^8\underline{L}$ final state (sharp peak), with a ligand hole, due to $p-d$ electron transfer. 
We use this picture to discuss our DMFT results.
A sizable $d^8\underline{L}$ resonant peak appears in the filled $t_{2g}$ band, 
in spite of only a weak $p-t_{2g}$ hybridization. This is interpreted as being due to
the emission of a $t_{2g}$ electron followed by an electron transfer from oxygen
to the partially filled $e_g$ shell.
The $e_g$ spectrum exhibits even richer structure.  
In the conduction sector we observe, besides the dominant $d^9$ peak, also
a tiny high frequency feature due to the $d^{10}\underline{L}$ final state, an accurate resolution of 
which is difficult using the maximum entropy method.
The the resonant $d^8\underline{L}$ peak exhibits a pronounced splitting in the $e_g$ channel.
We ascribe this to formation of the Zhang-Rice bound state
\cite{zan89} at the gap edge, which was studied in detail in Ref.
\onlinecite{bala}.
 Similar effect was also observed on a simpler two band p-d model
\cite{kun,zol99}.
In the present work an accurate resolution of the fine structure of
the resonant peak is not possible since the spin-flip terms in the
interaction are not included.
\begin{figure}
\includegraphics[angle=270,width=\columnwidth,clip]{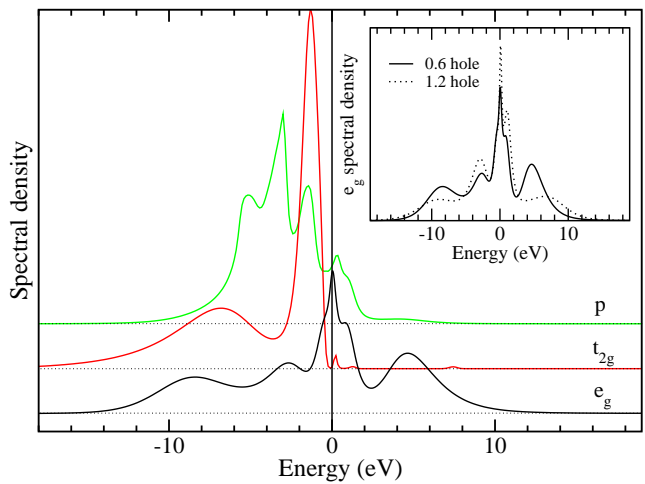}
\caption{\label{fig:dope2}(color online) Ni-$d$ and O-$p$ resolved spectral densities for a hole concentration
$n_h=0.6$ (offset for better resolution). The inset shows a comparison of $e_g$ spectral densities for hole
concentrations $n_h=0.6$ and $1.2$.}
\end{figure}

\subsection{Hole doping}
\begin{figure}
\includegraphics[angle=270,width=\columnwidth,clip]{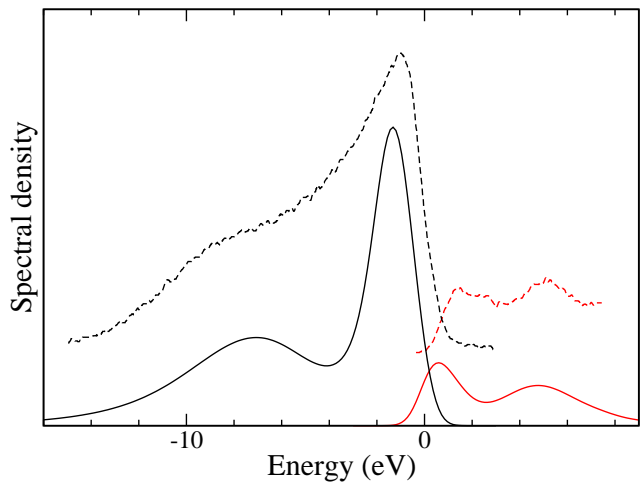}
\caption{\label{fig:linio2}(color online) Theoretical Ni-$d$ spectral densities for electron addition and
electron removal obtained for $n_h=0.6$ hole-doped NiO compared to the photoemission
and inverse photoemission spectra \cite{linio} of Li$_{0.4}$Ni$_{0.6}$O (the experimental
baseline is offset for better readability).}
\end{figure}

Next we discuss hole doping of NiO. An experimental realization can be found in 
Li$_x$Ni$_{1-x}$O studied in the doping
range $x$=0.02-0.4 \cite{linio}. 
Using the Hamiltonian of the undoped system
the replacement of $x$ Ni$^{2+}$ ions by
Li$^{1+}$ ions introduces on average $n_h=x/(1-x)$ hole per Ni site. 
As crude as this approximation may be we believe that the essential
physics of p-d weight transfer is captured correctly.
In Fig. \ref{fig:dope2} we show
the single-particle spectral densities for $n_h=0.6$ corresponding to $x=0.38$. There is no significant
difference between the $t_{2g}$ spectra in the doped and undoped cases, but the $e_g$ spectral function changes 
significantly. Most notably the Mott gap is filled, while the Hubbard subbands are preserved as distinct features.
This is also observed in experiment, as shown in Fig. \ref{fig:linio2}.  
A quite different behavior was reported in the single-band Mott insulator \cite{pru93}, where
the gap survives doping while the coherent peak merges with one of the Hubbard subbands. 
Further hole doping of NiO leads to a spectral weight transfer
from both upper- and lower-Hubbard-subbands to the quasiparticle part of the $e_g$ spectrum
(see inset of Fig. \ref{fig:dope2}), which can be viewed as an enhancement 
the itinerant character of the system reflected also in the decrease of the local moment (see Table \ref{tab:1}).
The $p$ spectral density suggests that much of the doped hole
is accommodated on the oxygen sites. That this is indeed the case can be seen from
the reduction of the $p$ orbital occupancy shown in Table \ref{tab:1}. 
In an uncorrelated system doping results in a mere shift of the chemical potential 
and the variation of the orbital occupancy is given by
the spectral density at the chemical potential. 
This reasoning clearly fails in a multi-orbital correlated system such as NiO.
While the Ni-$d$ spectral density 
at the top of the valence band in the stoichiometric system (or at the chemical potential
in the doped system) is comparable to the O-$p$ spectral density, doped holes almost exclusively
reside at the oxygen sites. 

\section{Conclusions}
We used the LDA+DMFT(QMC) approach to compute single-particle spectra of the prototypical CT insulator 
NiO. By including the ligand $p$ states and the on-site Coulomb interaction within the same framework we were able 
provide a full description of the valence band spectrum and, in particular, of the distribution
of spectral weight between the lower Hubbard band and the resonant peak at the top of the valence band. 
Good agreement with the available photoemission and inverse-photoemission data was found without need
for adjustable parameters. Importantly, the present method allows us to study the hole doped regime
where we find that the doped holes are distributed mainly among the ligand sites. 
A high hole doping leads to the filling of the correlation gap and a significant
transfer of the $d$-spectral weight form the incoherent part of the spectrum.

\acknowledgements
We thank W. E. Pickett and R. T. Scalettar for numerous discussions at the early stage of the code development.
J.K. was sponsored by a Research Fellowship of the Alexander von Humboldt Foundation.
J.K. and D.V. acknowledge partial support by the SFB 484 of the Deutsche Forschungsgemeinschaft. 
V.I.A. and A.V.L were supported by the Russian Foundation for Basic Research
under the grants RFFI-06-02-81017, RFFI-04-02-16096, and RFFI-03-02-39024,
and by the Netherlands Organization for Scientific Research through NWO 047.016.005.
A.V.L acknowledges support from the Dynasty Foundation and International 
Center.

\end{document}